# Enhanced *d–p* hybridization intertwined with anomalous ground state formation in van der Waals-coupled magnetic metal Fe$_5$GeTe$_2$

K. Yamagami[1], Y. Fujisawa[1], M. Pardo-Almanza[1], B. R. M. Smith[1], K. Sumida[2], Y. Takeda[2], and Y. Okada[1]

[1]*Okinawa Institute of Science and Technology (OIST) Graduate University,*

*Tancha, Onna-son, Kunigami-gun Okinawa, 904-0495, Japan.*

[2] *Materials Sciences Research Center, Japan Atomic Energy Agency (JAEA),*

*Sayo-cho, Sayo-gun, Hyogo 679-5148, Japan.*

**Abstract**

Fe$_5$GeTe$_2$ is a van der Waals (vdW)-coupled unconventional ferromagnetic metal with a high Curie temperature ($T_C$) exceeding 300 K. The formation of an anomalous ground state significantly below $T_C$ has received considerable attention, resulting in increased interest in understanding the spin-polarized electronic state evolution near the Fermi energy ($E_F$) as a function of temperature. Despite recent extensive studies, a microscopic understanding of the spin-polarized electronic structure around $E_F$ has not yet been established owing to the intrinsic complexity of both the crystal and band structures. In this study, we investigate the temperature dependence of element-specific soft X-ray magnetic circular dichroism (XMCD). A systematic temperature evolution in the XMCD signal from both magnetic Fe and its ligand Te is clearly observed. More importantly, the enhancement in the hybridization between the Fe 3*d* and Te 5*p* states in the zero-magnetic field limit is revealed, and we discuss its implications on the possible emergence of an exotic magnetic ground state in Fe$_5$GeTe$_2$.

Materials hosting the coexistence of electronic conduction and magnetic order have been a crucial platform for identifying and investigating exotic correlated phases [1,2,3]. Magnetism typically originates from partially filled and relatively localized $d$ or $f$ electron orbitals in transition metals or rare-earth elements. By controlling their hybridization with delocalized ligand $s$ and/or $p$ orbitals, the physics of correlated systems can be enriched [1,2,4,5,6,7,8]. One of the most exotic cases is cuprates, in which the control of hybridization between Cu $3d$ and oxygen $2p$ orbitals ($d$–$p$ hybridization) is crucial for controlling the competition of multiple exotic phases [9,10]. Heavy fermion (HF) systems are another good example, in which hybridization between $f$ electron orbitals from rare-earth elements and conduction electrons (often known as $c$–$f$ hybridization) are crucial for tuning the system from a magnetically ordered state to an HF state [4,5,6,7,8]. The recent surge in interest in identifying exotic metallic states in van der Waals (vdW)-coupled magnetic systems [11,12,13] necessitates an understanding and control of the hybridization between magnetic-localized and non-magnetic delocalized electronic orbitals. However, such an understanding has not been well developed, partly due to the limited number of vdW materials hosting both the electronic conductivity and magnetic order.

Recently, the Fe$_n$GeTe$_2$ ($n$ = 3, 4, 5) family has received considerable interest as a promising vdW material series for hosting the coexistence of electronic conduction and magnetic ordering [14,15,16,17,18,19,20,21,22,23,24,25,26,27]. Within this group, various topologically nontrivial states and consequent exotic transport phenomena have been observed [28]. Fe$_5$GeTe$_2$ is of particular interest owing to its high Curie temperature ($T_C$) [see Fig. 1(a) for the structure]. The Fe$_5$GeTe$_2$ system has been recognized to host unconventional ferromagnetism [19,20], whose microscopic spin structure can be better understood as antiferromagnetism or ferrimagnetism [29]. In addition to the topologically nontrivial magnetic state [28], the formation of various intriguing ground states has been argued for Fe$_5$GeTe$_2$ and its related compounds. For example, charge density wave (CDW) formation [25,30] and HF formation [22,31] have been discussed for this system below ~150 K (which is significantly below $T_C$). These observations highlight the importance of the hybridization degree, as in HF systems [4,5,6,7,8]. However, understanding hybridization, which should be reflected in the electronic structure near the Fermi level ($E_F$), is challenging, even though the intrinsically cleavable nature of this crystal should facilitate surface-sensitive high-resolution single-particle spectroscopy. In particular, the challenge is attributed to the intrinsic chemical complexity of Fe$_5$GeTe$_2$ [32,33,34,35]. It is known that 50% of Fe(1) sites are vacant even for the stoichiometric composition Fe$_5$GeTe$_2$ [see the circle marked by the partially occupied region denoted in color for Fe(1) site in Fig. 1(a)], in addition to the fact that the material tends to be invariably off-stoichiometrically deficient in Fe (e.g., Fe$_{5-x}$GeTe$_2$) [19,20,23]. The structural freedom of the Fe atoms occupying the Fe(1) sites inherently complicates the band structure, which prevents determining the exact hybridization degree using angle-resolved photoemission spectroscopy [30]. Hence, an alternative probe is required to determine the hybridization strength of the entire system, without considering the complex band structure.

X-ray magnetic circular dichroism (XMCD) measurements have recently been demonstrated as effective for understanding the complexity of Fe$_5$GeTe$_2$ [34]. XMCD measurements provide information regarding element-specific (orbital-selective) spin-polarized electronic density of states $N(\uparrow_{major}) - N(\downarrow_{minor})$, including those in the vicinity of $E_F$, as a function of the external magnetic field. Therefore, element-specific magnetization curves were obtained. It was demonstrated that the XMCD-based magnetization curve shapes for Fe $3d$ and Te $5p$ were consistent with the bulk magnetometry curves [Fig. 1(b)], while considering the fact that the orbital-dependent excitation process in the XMCD measurement merely reverses the sign of the XMCD signal [34]. This observation was interpreted as a consequence of hybridization-induced magnetism in the Te $5p$ states, since the localized Fe $3d$ orbital and delocalized ligand Te 5$p$ orbital are expected to coexist near $E_F$ [Fig. 1(c)] (see also Ref. [34]). Therefore, the degree of $d$–$p$ hybridization near $E_F$ can be probed sensitively via element-specific XMCD measurements in complex systems such as Fe$_5$GeTe$_2$. The previous XMCD study was performed only at 20 K, and a temperature-dependent study is yet to be conducted. Herein, we present the first study pertaining to the temperature and magnetic field evolution of element-specific XMCD in Fe$_5$GeTe$_2$, where enhanced $p$–$d$ hybridization significant below $T_C$ is revealed anomalously. We discuss the interpretation of these element-specific XMCD data based on the possible formation of an exotic ground state at low temperatures in the low-field limit.

A single crystal was synthesized using a previously described method [19,20]. Based on X-ray diffraction (XRD), the absence of undesired crystalline phases was confirmed. Energy-dispersive X-ray (EDX) spectroscopy (Quanta 250 FEG, FEI) results show the actual composition of Fe$_{4.52}$Ge$_{1.02}$Te$_2$ ($x \approx 0.48$). For simplicity, we refer to our compound as Fe$_5$GeTe$_2$ hereinafter. Magnetometry measurements using an MPMS-3 (Quantum Design) revealed a $T_C$ of approximately 310 K and absence of excessive iron atoms between the layers is confirmed (see Supplemental information [14,19,21,23,29,36,37,38,39,40]). For an unbiased comparison with the XMCD data, the magnetometry data shown herein were measured based on a $\theta = 45°$ geometry (see Fig. 2(a)).

Soft X-ray absorption (XAS) measurements, including XMCD measurements, were performed at BL23SU of SPring-8, which was equipped with two twin-helical-undulators that produce nearly perfectly left and right circularly polarized X-rays [41]. For the cleaved single crystal in ultrahigh vacuum, all XAS spectra were captured by the total electron yield mode below the energy resolution of < 100 meV and then normalized by the incident photon flux. The XMCD spectra were obtained from $\mu^- - \mu^+$, where $\mu^+$ ($\mu^-$) denotes the XAS intensity corresponding to the parallel (antiparallel) orientations of the sample magnetization and incident photon helicity [Fig. 2(a)]. During data acquisition, the circular polarization at each photon energy was switched on at a frequency of 1 Hz using five kicker magnets, which efficiently acquired XMCD data with a high signal-to-noise ratio. To eliminate experimental errors, the XAS/XMCD spectra with positive and negative magnetic fields were measured and averaged for each photon energy. We measured the XMCD signal based on a $\theta = 45°$ geometry [see Fig. 2(a)] because the magnetic anisotropy energy was relatively low in this system [19,20,36]. To investigate the spin-polarized density of states for Fe 3$d$ and Te 5$p$ [see Fig. 2(b)], as in our previous study [34], we focused on the Fe $L_{2,3}$- and Te $M_{4,5}$-edges. The XMCD signal from Ge [34] is not discussed herein because of its low intensity.

The XAS and XMCD spectra were measured under 0.2 and 4 T at 20, 120, and 230 K. Figs. 2(c) and (d) show the XAS and XMCD spectra at 0.2 T and 20 K for the Fe $L_{2,3}$- and Te $M_{4,5}$-edges, respectively. Other XAS and XMCD spectra are provided in the Supplemental Information [36]. Consistent with previous measurements, the Fe $L_{2,3}$- and Te $M_{4,5}$-edges appeared in the energy ranges of 690–760 and 565–595 eV, respectively. As discussed in our previous paper [34], disentangling the signal from nonequivalent Fe atomic sites is extremely difficult. Therefore, the XMCD signal for Fe is regarded as the summation of the signals from all non-equivalent Fe sites [see Fig. 1(a)]. For convenience, we define the XMCD intensity at the Fe $L_3$(~707 eV)- and Te $M_5$(~572 eV)-edges as $I_{Fe}$ and $I_{Te}$, respectively [see colored horizontal arrows in Figs. 2(c) and 2(d)].

To confirm that the XMCD signal is intrinsic based on the bulk, bulk magnetization (M–H), $I_{Te}$–H, and $I_{Fe}$–H curves were compared. It is known that the sign of the XMCD signal is determined by the change in the orbital angular momentum $\Delta l$ (= $l_{final} - l_{initial}$) during the excitation process. Because the sign of the XMCD at the $L_{2,3}$ and $M_{4,5}$ edges corresponding to the $p \rightarrow d$ ($\Delta l = 1$) and $d \rightarrow p$ processes ($\Delta l = -1$) [see

Fig. 2(c) and 2(d)] is negative, we reversed the sign of the M–H curve to compare it with $I_{Fe}$ and $I_{Te}$. Comparisons of $I_{Fe}$ and $I_{Te}$ between +4 T and −4 T are shown in Figs. 2(e) and 2(f), respectively. Within our measurement resolution, all bulk magnetometry and XMCD curves increased linearly and saturated within ±1 T without indicating clear hysteresis. This behavior is similar to that observed for different sample configurations relative to the magnetic field, which suggests a slight magnetic anisotropy [34,36]. As shown in Figs. 2(e)–(f), qualitatively similar M–H, $I_{Fe}$–H, and $I_{Te}$–H curve shapes were indicated for all temperatures. This intimate connection between XMCD and bulk magnetism was further confirmed by comparing the temperature evolutions of $I_{Fe}$ and $I_{Te}$ [Fig. 3(a)]. As temperature decreased, both $I_{Fe}$ and $I_{Te}$ at 4 T showed a systematic increase. By contrast with both $I_{Fe}$ and $I_{Te}$ at 4 T, the opposite behavior at 20 K was observed at 0.2 T. This characteristic difference in the magnetic signal between 0.2 and 4 T in the XMCD measurement is consistent with the bulk magnetometry data [solid and dashed black curves in Fig. 3(a)]. This confirms that our elemental-selective XMCD results reflect the intrinsic nature of bulk electronic and the magnetic properties. A key advantage of element-specific XMCD, however, is its ability to extract unique information from bulk magnetometry.

To understand the element-specific nature of XMCD, we employed the relevant XMCD signal, $I_{Te}/I_{Fe}$. As shown in Fig. 3(a), $I_{Te}$ improved as compared with $I_{Fe}$ at 20 K and 0.2 T. The enhancement in $I_{Te}$ was further supported by a comparison of the magnetic field evolution of the $I_{Te}/I_{Fe}$ ratio among three temperatures of 20, 120, and 230 K [Fig. 3(b)]. Because no hysteresis component was present, we show the ratio only for the absolute field range after averaging the ratios for the positive and negative fields. In contrast to the 120 and 230 K cases, $I_{Te}/I_{Fe}$ enhanced significantly in the low H-field region (< 1 T) at 20 K. The increased $I_{Te}/I_{Fe}$ ratio is particularly interesting because a stronger spin polarization of the Te $5p$ state is suggested by the enhancement in hybridization of the Fe $3d$ state. As will be shown later, this is supported by magneto–optical sum-rule analysis [42,43,44].

Magneto–optical sum-rule analysis is performed to evaluate the relative magnetic moment ($m_{orb}/m_{spin}$) from the orbital ($m_{orb}$) and effective spin ($m_{spin}$) components of the target atom. As schematically shown in Fig. 4(a), intensifying the localized nature of the magnetic moment should result in a higher $m_{orb}/m_{spin}$ value, and intensifying the delocalized nature should result in a lower $m_{orb}/m_{spin}$ value. Therefore, $m_{orb}/m_{spin}$ can indicate the hybridization degree qualitatively, as demonstrated in doping-induced ferromagnetic systems [45] and HF systems [46]. In the case of $Fe_5GeTe_2$, because the Fe $3d$ and Te $5p$ electronic orbitals are localized and delocalized, respectively [see Fig. 1(c)], the temperature dependence of $m_{orb}/m_{spin}$ between Fe $3d$ and Te $5p$ is expected to exhibit opposite behaviors if the $d$–$p$ hybridization is enhanced. In this study, by focusing on XMCD in 0.2 T, we performed a magneto–optical sum-rule analysis for both the Fe $L_{2,3}$- and Te $M_{4,5}$-edges to estimate $m_{orb}/m_{spin}$. In fact, the $m_{orb}/m_{spin}$ for Fe can be calculated from the integration of the XMCD intensity for the overall $L_3$ edge (700–717 eV) and the overall $L_2 + L_3$ edge (700–750 eV) [Fig. 2(b)]. Similarly, the $m_{orb}/m_{spin}$ for Te can be calculated from the integration of the XMCD intensity for the overall $M_5$ edge (565–581 eV) and $M_4 + M_5$ edges (565–593 eV) [Fig. 2(c)]. Further details are available in the Supplementary Information [36,45,46,47,48].

Figure 4(b) shows the temperature dependence of $m_{orb}/m_{spin}$ for the Fe $3d$ and Te $5p$ states at 0.2 T. By performing cooling, the $m_{orb}/m_{spin}$ value of Fe decreased, whereas that of Te increased. This contrasting trend can be reasonably attributed to a counter-effect due to the increasing hybridization between delocalized Te $5p$ and localized Fe $3d$ orbitals [see Fig. 1(c)]. To further illustrate this contrasting trend, the ratio $(m_{orb}/m_{spin})_{Te}/(m_{orb}/m_{spin})_{Fe}$ was compared with $I_{Te}/I_{Fe}$ at 0.2 T [Fig. 4(c)]. The value of $(m_{orb}/m_{spin})_{Te}/(m_{orb}/m_{spin})_{Fe}$ reflects the enhancement degree of the localized nature of Te $5p$, and its excellent scaling with $I_{Te}/I_{Fe}$ led us to conclude that Fe $3d$–Te $5p$ hybridization increases prominently with decreasing temperature in the low magnetic field limit.

At the phenomenological level, we discuss the connection between enhanced $d$–$p$ hybridization and the anomalous ground-state formation significantly below $T_C$. One possible scenario is the existence of CDW ordering below ~150 K, as reported in recent publications [25,30]. The emergence of a CDW can cause band folding in the reciprocal lattice space owing to symmetry breaking by periodic lattice modulation, which is analogous to cases involving excitonic charge density wave materials $TiSe_2$ and $ZrTe_2$ [49,50]. For ferromagnetic metal $Fe_5GeTe_2$, because the relatively localized Fe $3d$ band and delocalized Te $5p$ band are hybridized [34], band folding due to CDW ordering results in more crossings between the Fe $3d$ and Te $5p$ bands, resulting in enhanced $d$–$p$ hybridization. Supporting the low-dimensional double exchange model [51], the enhanced $d$–$p$ hybridization can induce a relatively delocalized nature in the Fe $3d$ electrons [Figs. 4(a) and 4(b)] and the small but finite gap in the spin-polarized conduction band, which consequently weakens the ferromagnetism. Another possible scenario is the existence of an HF state, as proposed for $Fe_3GeTe_2$ [22,31]. The magnetic moment from Fe $3d$ electrons is screened by the spins of delocalized Te $5p$ electrons owing to the antiferromagnetic spin arrangement between localized Fe $3d$ and delocalized Te $5p$ electrons [34]. A $d$–$p$ hybridized HF band is formed near $E_F$ at low temperatures, consequently resulting in the suppression of ferromagnetism, as in HF systems [7,8]. Our temperature-dependent XMCD results [see Supplemental information [36]] do not contradict both the CDW and HF scenarios below 150 K. If these CDW and/or HF formation scenarios are in fact correct, then the melting of the CDW and/or Kondo lattice by applying an external field above ~1 T is expected, based on the magnetic field evolution of $I_{Te}/I_{Fe}$ at 20 K [see blue curve in Fig. 3(b)].

In summary, we demonstrated the temperature and magnetic field dependence of element-specific XMCD measurements in a $Fe_5GeTe_2$ system. Based on a cooled sample, a systematic increase in the magnetic moment of the Te $5p$ state due to enhanced $3d$–$5p$ hybridization was clarified. Whereas the ligand state has been regarded as a less significant component compared with the Fe $3d$ orbitals, the enhancement of its hybridization with the 5p orbital in heavy element Te should not be underestimated for modeling the ground state magnetic background, and hence its excited states. In particular, the significant spin–orbit coupling effect from heavy element Te might modify the magnetic crystalline and exchange anisotropies. Because an intrinsic large tunability of occupation exists at the Fe(1) site, which is the nearest Fe site to the Te atomic sheet, our findings provide important information for identifying exotic magnetic ground states and their excited states in $Fe_5GeTe_2$. We believe that the findings presented herein would facilitate the deeper understanding of itinerant magnetism in low-dimensional systems, as many itinerant magnetic chalcogenides contain heavy Te. Furthermore, in this study, XMCD measurements were clearly demonstrated as one of the most effective methods for clarifying near-$E_F$ element-specific spin-polarized electronic states, particularly for materials with intrinsically complicated band structures.

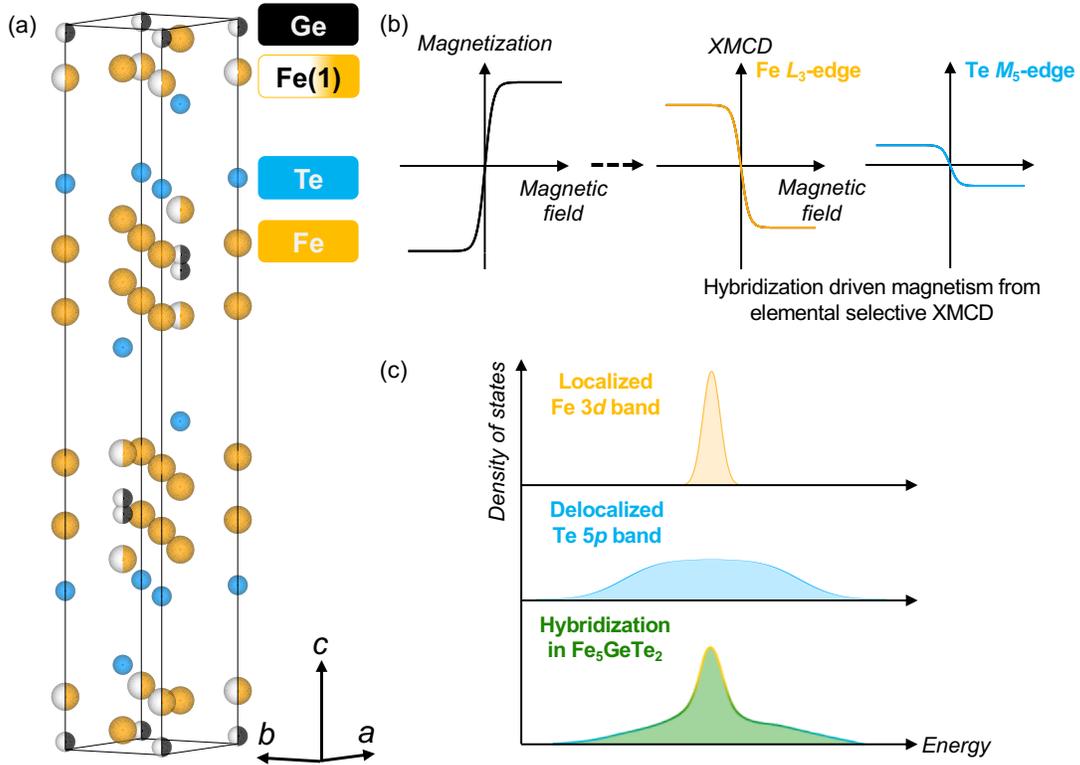

**Figure 1.** (Color online) (a) Crystal structure of $Fe_5GeTe_2$. Solid black lines correspond to unit cell. In $Fe_5GeTe_2$, 50% split occupation for Fe(1) and Ge sites is known [19,20,23]. (b) Schematic illustration showing elemental-selective magnetization curves for Fe 3$d$ (yellow) and Te 5$p$ (light blue) states measured via XMCD for $Fe_5GeTe_2$, which constitute the bulk magnetization curve (black). XMCD signal from Te is associated with a hybridization-induced origin, as per literature [34]. Sign of XMCD signal is opposite compared with bulk magnetization curves owing to definition of obtained XMCD spectra in circular-polarized X-ray absorption measurement under external magnetic field (see main text for details). (c) Schematic illustration of localized Fe 3$d$ (yellow) and delocalized Te 5$p$ (light blue) states, together with hybridized states between those orbitals (green), as obtained from literature [34]. When the hybridization was weak, the bandwidths of Fe 3$d$ and Te 5$p$ were narrow and wide, respectively. By contrast, when the hybridization became stronger, they the band properties of both were included, and Fe 3$d$ and Te 5$p$ became broader and narrower, respectively. Herein, we refer to this relationship as the counter-effect.

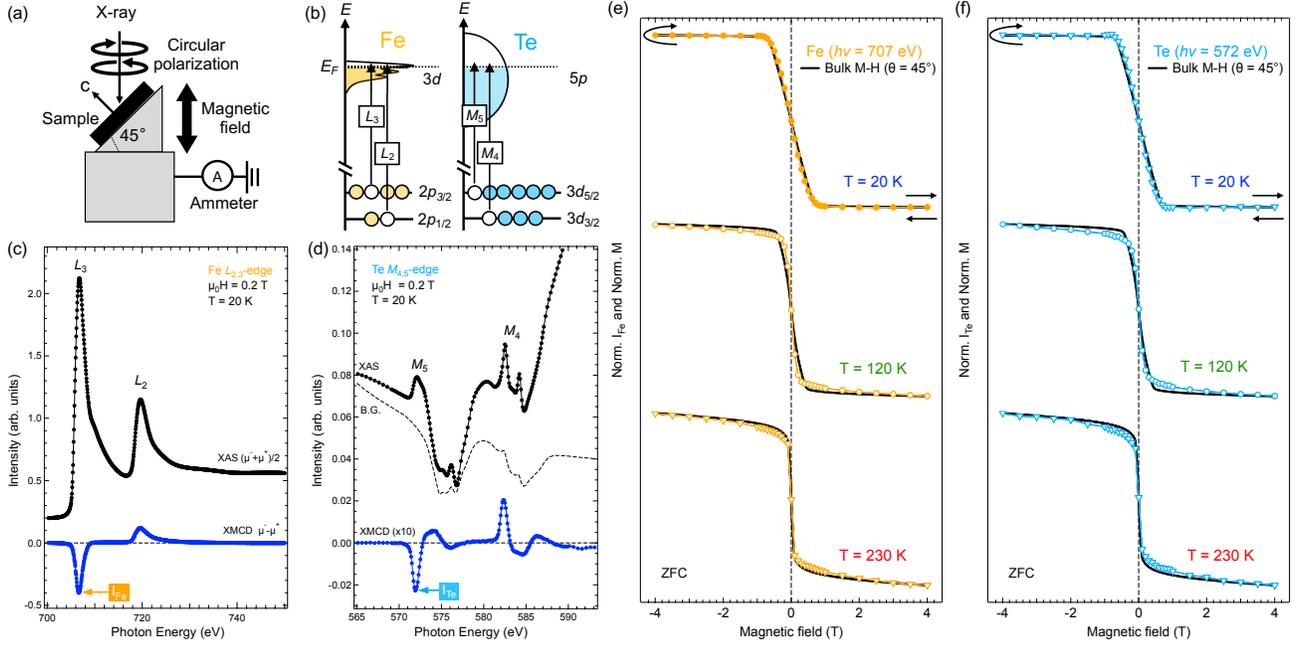

**Figure 2.** (Color online) (a) Geometry of XMCD measurements in total electron yield (TEY) mode with $\theta = 45°$. XMCD spectra are defined as $\mu^- - \mu^+$, where $\mu^+$ ($\mu^-$) denotes absorption intensity for parallel (antiparallel) alignment of photon helicity and sample magnetization direction. (b) Schematic illustration of excitation process for Fe $L_{2,3}$- and Te $M_{4,5}$-edges. (c) Te $L_{2,3}$-edge and (d) Te $M_{4,5}$-edge XMCD spectra under 0.2 T at 20 K. Definition of characteristic intensities $I_{Fe}$ and $I_{Te}$ at $L_3$ (~707 eV) and $M_5$ (~572 eV) are indicated by horizontal arrows (see main text for details). In Te $M_{4,5}$ edge, the background spectrum including in the Cr $L_{2,3}$ edge ($\approx$ 576 and 585 eV) absorption from the focusing mirrors in the BL23SU optical system are shown as the black solid line [34]. (e) – (f) Comparison of magnetization curve shape between bulk M–H curves (with angle of induced magnetic field to sample surface normal direction $\theta$ of 45°), $I_{Fe}$–H curves, and $I_{Te}$–H curves. All data presented were obtained under zero-field cooling (ZFC) conditions and measured by changing the magnetic field, as indicated by the arrows. Within the resolution in our XMCD measurements, no hysteresis component was observed. In (e) and (f), the sign of the M–H curves are reversed (see main text for details).

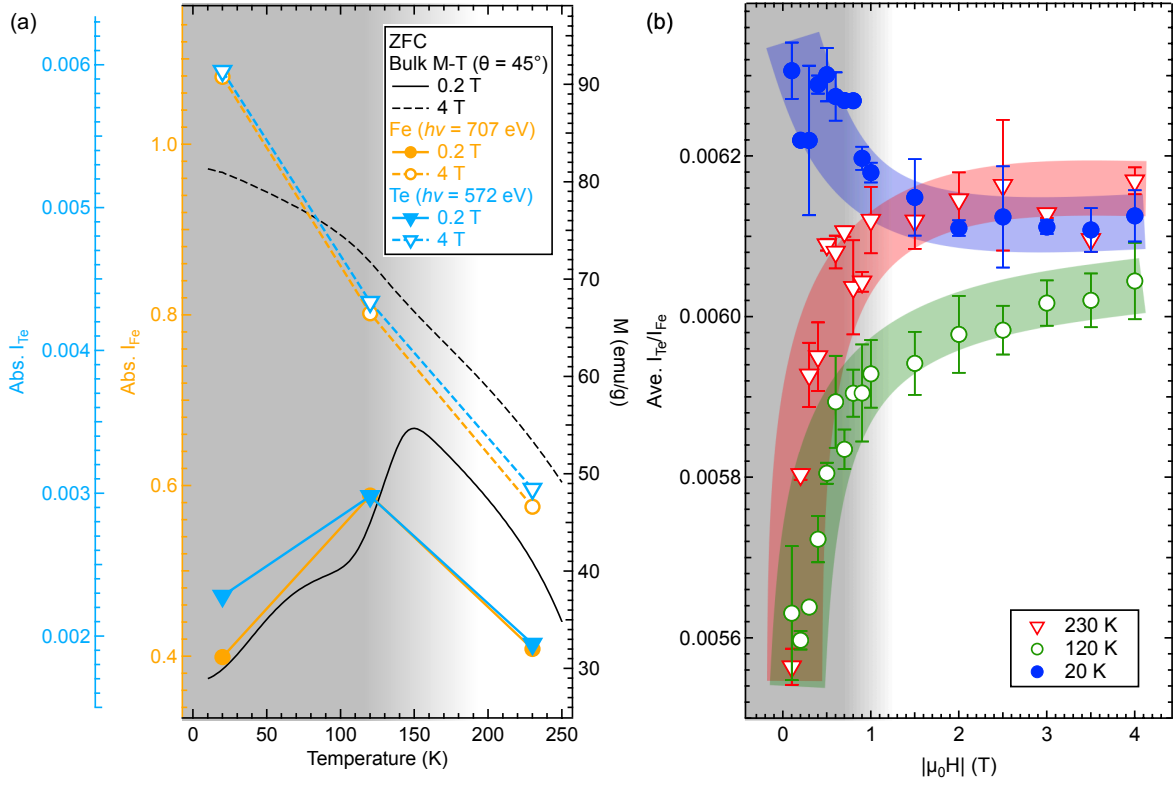

**Figure 3.** (Color online) (a) Temperature dependence of $I_{Fe}$ and $I_{Te}$ at 0.2 T and 4 T. Here, $I_{Fe}$ and $I_{Te}$ are defined as the XMCD intensity at Fe $L_3$ (~707 eV) and Te $M_5$ (~572 eV), respectively [see Figs. 2(c) and (d)]. M–T curves (ZFC) from macroscopic magnetometry with $\theta = 45°$ are shown as solid and dashed black lines. (b) Magnetic field dependence of averaged ratio $I_{Te}/I_{Fe}$ at 20, 120, and 230 K, as obtained from curves shown in Figs. 2(e) and (f). As no hysteresis component was observed, the averaged ratio $I_{Te}/I_{Fe}$ for positive and negative fields was plotted as a function of the absolute value of the external magnetic field ($|\mu_0 H|$). Error bar denotes variation in ratio between positive and negative fields. Data at $\mu_0 H = 0$ T was removed to avoid mathematical singular point.

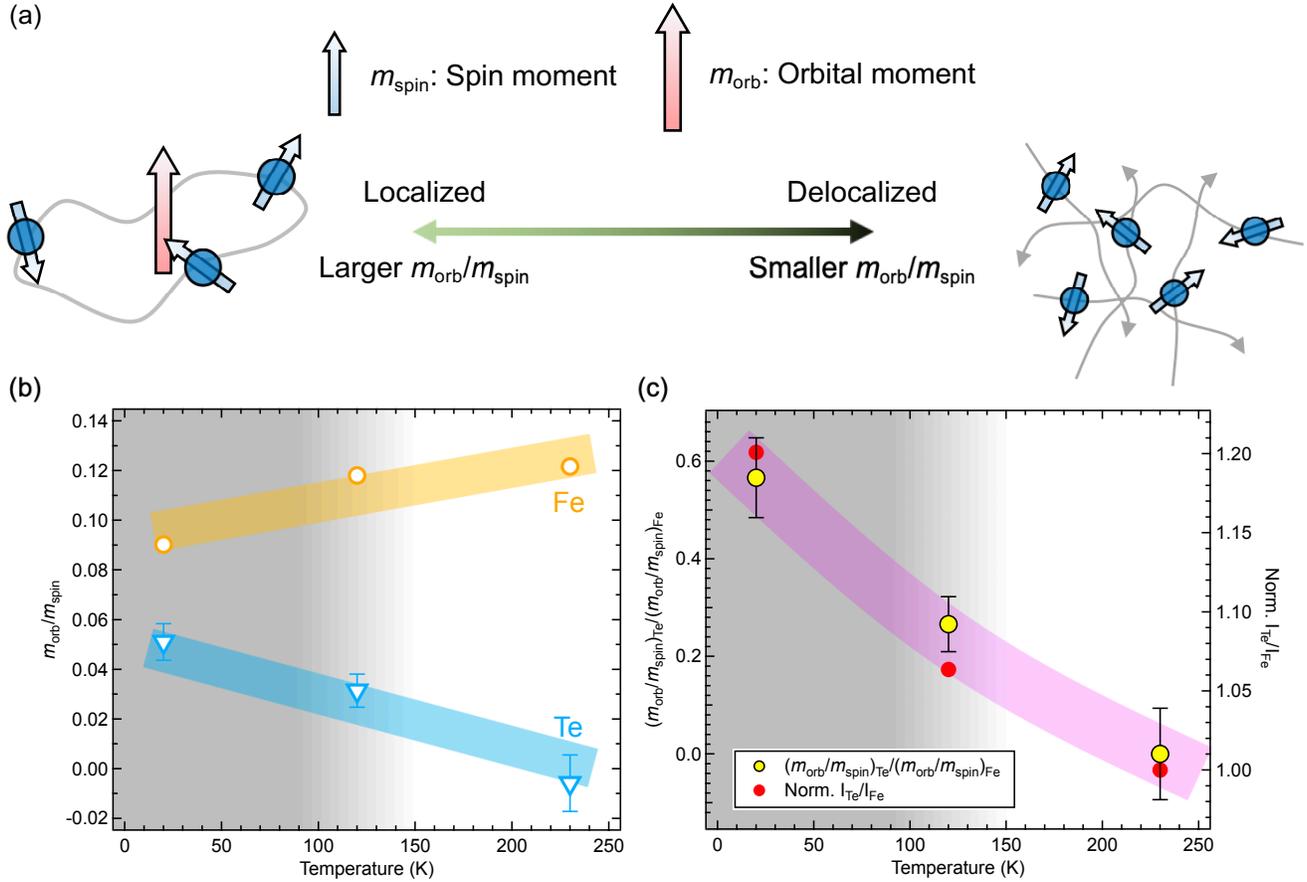

**Figure 4.** (Color online) (a) Schematic illustration of momentum from orbital ($m_{orb}$) and spin ($m_{spin}$) components (see main text for details). (b) Comparison of temperature-dependent $m_{orb}/m_{spin}$ between Fe and Te at 0.2 T. Opposite temperature dependence for Fe and Te are highlighted by lines. (c) Comparison between $(m_{orb}/m_{spin})_{Te}/(m_{orb}/m_{spin})_{Fe}$ and $I_{Te}/I_{Fe}$ at 0.2 T. Intimate scaling was observed between these two values (see main text for details). Details regarding estimation of error bar in panels (b) and (c) are provided in Supplemental Information [36].

## REFERENCES

[1] M. Imada, A. Fujimori and Y. Tokura, Rev. Mod. Phys. **70**, 1039 (1998).

[2] Y. Tokura and N. Nagaosa, Science **288**, 462 (2000).

[3] E. Dagotto, Science **309**, 257 (2005).

[4] C. M. Varma, Rev. Mod. Phys. **48**, 219 (1976).

[5] J. M. Lawrence, P. S. Riseborough, and R. D. Parks, Rep. Prog. Phys. **44**, 1 (1981).

[6] G. R. Stewart, Rev. Mod. Phys., **56**, 755 (1984).

[7] S. Doniach Physica B + C **91**, 231 (1977).

[8] Z. F. Weng, M. Smidman, L. Jiao, X. Lu, and H. Q. Yuan, Rep. Prog. Phys. **79**, 094503 (2016).

[9] F. C. Zhang and T. M. Rice Phys. Rev. B **37**, 3759 (1988).

[10] B. Keimer, S. A. Kivelson, M. R. Norman, S. Uchida, and J. Zaanen, Nature **518**, 179 (2015).

[11] A. K. Geim and I. V. Grigorieva, Nature **499**, 419 (2013).

[12] C. Gong and X. Zhang, X. Science **363**, 706 (2019).

[13] S. Yang, T. Zhang, and C. Jiang, Adv. Sci. **8**, 2002488 (2021).

[14] B. Chen, J.H. Yang, H.D. Wang, M. Imai, H. Ohta, C. Michioka, K. Yoshimura, and M.H. Fang, J. Phys. Soc. Jpn. **82**, 124711 (2013).

[15] Z. Fei, B. Huang, P. Malinowski, W. Wang, T. Song, J. Sanchez, W. Yao, D. Xiao, X. Zhu, A. F. May, W. Wu, D. H. Cobden, J.-H. Chu, and X. Xu, Nat. Mater. **17**, 778 (2018).

[16] G. D. Nguyen, J. Lee, T. Berlijn, Q. Zou, S. M. Hus, J. Park, Z. Gai, C. Lee, and A.-P. Li, Phys. Rev. B **97**, 014425 (2018).

[17] Y. Deng, Y. Yu, Y. Song, J. Zhang, N. Z. Wang, Z. Sun, Y. Yi, Y. Z. Wu, S. Wu, J. Zhu, J. Wang, X. H. Chen, and Y. Zhang, Nature (London) **563**, 94 (2018).

[18] Y.-P. Wang, X.-Y. Chen, and M.-Q. Long, Appl. Phys. Lett. **116**, 092404 (2020).

[19] A. F. May, D. Ovchinnikov, Q. Zheng, R. Hermann, S. Calder, B. Huang, Z. Fei, Y. Liu, X. Xu, and M. A. McGuire, ACS Nano **13**, 4436 (2019).

[20] A. F. May, C. A. Bridges, and M. A. McGuire, Phys. Rev. Mater. **3**, 104401 (2019).

[21] J. Seo, D. Y. Kim, E. S. An, K. Kim, G.-Y. Kim, S.-Y. Hwang, D. W. Kim, B. G. Jang, H. Kim, G. Eom, S. Y. Seo, R. Stania, M. Muntwiler, J. Lee, K. Watanabe, T. Taniguchi, Y. J. Jo, J. Lee, B. Il Min, M. H. Jo, H. W. Yeom, S.-Y. Choi, J. H. Shim, and J. S. Kim, Sci. Adv. **6**, eaay8912 (2020).

[22] Y. Zhang, H. Lu, X. Zhu, S. Tan, W. Feng, Q. Liu, W. Zhang, Q. Chen, Y. Liu, X. Luo, D. Xie, L. Luo, Z. Zhang and X. Lai, Sci. Adv. **4**, eaao6791 (2018).

[23] H. Zhang, R. Chen, K. Zhai, X. Chen, L. Caretta, X. Huang, R. V. Chopdekar, J. Cao, J. Sun, J. Yao, R. Birgeneau, and R. Ramesh, Phys. Rev. B **102**, 064417 (2020).

[24] J. Seo, E. S. An, T. Park, S.-Y. Hwang, G.-Y. Kim, K. Song, W.-S. Noh, J. Y. Kim, G. S. Choi, M. Choi, E. Oh, K. Watanabe, T. Taniguchi, J.-H. Park, Y. J. Jo, H. W. Yeom, S.-Y. Choi, J. H. Shim, and J. S. Kim, Nat. Commun. **12**, 2844 (2021).

[25] Y. Gao, Q. Yin, Q. Wang, Z. Li, J. Cai, T. Zhao, H. Lei, S. Wang, Y. Zhang, and B. Shen, Adv. Mater. **32**, 2005228 (2020).

[26] B. Ding, Z. Li, G. Xu, H. Li, Z. Hou, E. Liu, X. Xi, F. Xu, Y. Yao, and W. Wang, Nano Lett. **20**, 868 (2020).

[27] T J. Kim, S. Ryee, M. J. Han, arXiv:2202.02022v1

[28] K. Kim, J. Seo, E. Lee, K.-T. Ko, B. S. Kim, B. G. Jang, J. M. Ok, J. Lee, Y. J. Jo, W. Kang, J. H. Shim, C. Kim, H. W. Yeom, B. I. Min, B.-J. Yang, and J. S. Kim, Nat. Mater. **17**, 794 (2018).

[29] T. Ohta, K. Sakai, H. Taniguchi, B. Driesen, Y. Okada, K. Kobayashi, and N. Niimi, Appl. Phys. Express **13**, 043005 (2020).

[30] X. Wu, L. Lei, Q. Yin, N.-N. Zhao, M. Li, Z. Wang, Q. Liu, W. Song, H. Ma, P. Ding, K. Liu, Z. Cheng, H. Lei, and S. Wang, Phys. Rev. B **104**, 165101 (2021).

[31] M. Zhao, B.-B. Chen, Y. Xi, Y. Zhao, H. Xu, H. Zhang, N. Cheng, H. Feng, J. Zhuang, F. Pan, X. Xu, W. Hao, W. Li, S. Zhou, S. X. Dou, and Y. Du, Nano Lett. **21**, 6117 (2021).

[32] M. Joea, U. Yang and C. Lee, Nano Mater. Sci. **1**, 299 (2019).

[33] T. T. Ly, J. Park, K. Kim, H.-B. Ahn, N. J. Lee, K. Kim, T.-E. Park, G. Duvjir, N. H. Lam, K. Jang, C.-Y. You, Y. Jo, S. K. Kim, C. Lee, S. Kim, and J. Kim, Adv. Funct. Mater. 2009758 (2021).

[34] K. Yamagami, Y. Fujisawa, B. Driesen, C. H. Hsu, K. Kawaguchi, H. Tanaka, T. Kondo, Y. Zhang, H. Wadati, K. Araki, T. Takeda, Y. Takeda, T. Muro, F. C. Chuang, Y. Niimi, K. Kuroda, M. Kobayashi, and Y. Okada, Phys. Rev. B **103**, L060403 (2021).

[35] X. Yang, X. Zhou, W. Feng, and Y. Yao, Phys. Rev. B **104**, 104427 (2021).

[36] See Supplemental Material at xx for the detail of the magnetic characterization, the temperature dependence of XAS and XMCD spectra, and the magneto-optical sum-rules analysis for Fe $L_{2,3}$ and Te $M_{4,5}$-edges.

[37] M. E. Fisher, Rep. Prog. Phys. **30**, 615 (1967).

[38] Y. Liu, V. N. Ivanovski, and C. Petrovic, Phys. Rev. B **96**, 144429 (2017).

[39] J. M. Lu, O. Zheliuki, I. Leermarkers, N. F. Q. Yuan, U. Zeitler, K. T. Lawand, and J. T. Ye, Science **350**, 1353 (2015).

[40] Z. Li, W. Xia, H. Su, Z. Yu, Y. Fu, L. chen, X. Wang, N. Yu, Z. Zou, and Y. Guo, Sci. Rep. **10**, 15345 (2020).

[41] Y. Saitoh, Y. Fukuda, Y. Takeda, H. Yamagami, S. Takahashi, Y. Asano, T. Hara, K. Shirasawa, M. Takeuchi, T. Tanaka, and H. Kitamura, J. Synchrotron Radiat. **19**, 388 (2012).

[42] B. T. Thole, P. Carra, F. Sette, and G. van der Laan, Phys. Rev. Lett. **68**, 1943 (1992).

[43] P. Carra, B. T. Thole, M. Altarelli, and X. Wang, Phys. Rev. Lett. **70**, 694 (1993).

[44] C. T. Chen, Y. U. Idzerda, H.-J. Lin, N. V. Smith, G. Meigs, E. Chaban, G. H. Ho, E. Pellegrin, and F. Sette, Phys. Rev. Lett. **75**, 152 (1995).

[45] L. R. Shelford, T. Hesjedal, L. Collins-McIntyre, S. S. Dhesi, F. Maccherozzi, and G. van der Laan, Phys. Rev. B **86**, 081304(R) (2012).

[46] T. Okane, Y. Takeda, H. Yamagami, A. Fujimori, Y. Matsumoto, N. Kimura, T. Komatsubara, and H. Aoki, Phys. Rev. B **86**, 125138 (2012).

[47] K. Takubo, R. Comin, D. Ootsuki, T. Mizokawa, H. Wadati, Y. Takahashi, G. Shibata, A. Fujimori, R. Sutarto, F. He, S. Pyon, K. Kudo, M. Nohara, G. Levy, I. S. Elfimov, G. A. Sawatzky, and A. Damascelli, Phys, Rev. B **90**, 081104(R) (2014).

[48] M. Ye, T. Xu, G. Li, S. Qiao, Y. Takeda, Y. Saitoh, S.-Y. Zhu, M. Nurmamat, K. Sumida, Y. Ishida, S. Shin, and A. Kimura, Phys. Rev. B **99**, 144413 (2019).

[49] H. Cercellier, C. Monney, F. Clerc, C. Battaglia, L. Despont, M. G. Garnier, H. Beck, P. Aebi, L. Patthey, H. Berger, and L. Forró Phys. Rev. Lett. **99**, 146403 (2007)

[50] Y. Song, C. Jia, H. Xiong, B. Wang, Z. Jiang, K. Huang, J. Hwang, Z. Li, C. Hwang, Z. Liu, D. Shen, J. Sobota, P. Kirchmann, J. Xue, T. P. Devereaux, S.-K. Mo, Z.-X. Shen, and S. Tang, arXiv:2201.11592 (2022).

[51] S. Nishimoto and Y. Ohta, Phys. Rev. Lett. **109**, 076401 (2012).

[52] K. Momma and F. Izumi, J. Appl. Crystallogr. **44**, 1272 (2011).